\documentclass{article}
\usepackage[a4paper]{geometry}

\usepackage{amsmath,amssymb,amsfonts}
\usepackage{graphicx}
\usepackage{hyperref}

\usepackage{authblk}
\usepackage{bm}
\usepackage{booktabs}
\usepackage[capitalise]{cleveref}
\usepackage[inline]{enumitem}
\usepackage{mathtools}
\usepackage{physics}
\usepackage{tikz}
\usepackage{siunitx}

\usetikzlibrary{arrows.meta, backgrounds, fit, calc}

\renewcommand{\vec}{\bm}
\newcommand{\mat}{\bm}

\newcommand{\T}{\intercal}
\newcommand{\xa}{\ensuremath{n}} % r
\newcommand{\xb}{\ensuremath{w}} % v
\newcommand{\y}{\ensuremath{a_{e,\mathrm{lim}}}}
\newcommand{\w}{\ensuremath{\vec \theta}} % weight vector of SVM

\newcommand{\CNOT}{\mathbf{CNOT}}

\DeclareMathOperator{\sign}{sign}

\title{Predicting Machining Stability\\with a Quantum Regression Model}
\author[1]{Sascha M\"ucke}
\author[1,2]{Felix Finkeldey}
\author[3]{Nico Piatkowski}
\author[2]{Tobias Siebrecht}
\author[1,2]{Petra Wiederkehr}
\affil[1]{Lamarr Institute, TU Dortmund University, Dortmund, Germany (\texttt{\{first\}.\{last\}@tu-dortmund.de)}}
\affil[2]{Virtual Machining, TU Dortmund University, Dortmund Germany (\texttt{\{first\}.\{last\}@tu-dortmund.de)}}
\affil[3]{Fraunhofer IAIS, Sankt Augustin, Germany (\texttt{nico.piatkowski@iais.fraunhofer.de)}}

\begin{document}

\maketitle

\begin{abstract}
In this article, we propose a novel quantum regression model by extending the Real-Part Quantum SVM.
We apply our model to the problem of stability limit prediction in milling processes, a key component in high-precision manufacturing.
To train our model, we use a custom data set acquired by an extensive series of milling experiments using different spindle speeds, enhanced with a custom feature map.
We show that the resulting model predicts the stability limits observed in our physical setup accurately, demonstrating that quantum computing is capable of deploying ML models for real-world applications.
\end{abstract}

\section{Introduction}
\label{sec:introduction}

In the light of rapid developments in quantum hardware, the field of quantum computing (QC) has regained substantial interest over the past decade.
Particularly the field of quantum machine learning (QML) has emerged as the intersection of QC and classical machine learning (ML), where researchers try to harness the manipulation and measurement of quantum states for performing classification and regression tasks.
A substantial portion of research focuses on designing heuristic approaches, while the true expressive capabilities of quantum models and how to apply them deliberately is still a widely disregarded and poorly understood topic, even though some attempts at a rigorous formalization exist~\cite{schuld2021quantum,schuld2021effect}.

One prominent example of a largely heuristic approach to QML is the Quantum Support Vector Machine (QSVM)~\cite{havlivcek2019supervised}, which consists of a parametrized quantum circuit with a fixed gate structure, known as ``ansatz'' in literature, where the data and parameter values are given as input, the resulting quantum state is measured, and the obtained binary vectors are mapped to binary labels, e.g., using a parity function.
The parameters are trained in a variational fashion to minimize some loss function.
While it can be shown that in certain orthogonal Hermitian matrix bases, such QSVM decision functions resemble that of the classical SVM, such a basis transformation is not given explicitly, and there is no structural risk minimization such as margin maximization, which is the hallmark of the classical SVM.
Therefore, the connection between QSVMs and their classical counterpart underpinned by foundational learning theory is tenuous.

In this article, we build upon the recently proposed Real-Part Quantum SVM (RQSVM) model~\cite{piatkowski.muecke.2024a}, which is a quantum model that maintains the theoretical properties of the classical SVM, replicating its behavior exactly when the number of quantum measurements approaches infinity.
We extend this model by turning it from a classification into a regression model, and apply it to real-world data to show its effectiveness.

Data taken from real-world industrial applications is often challenging for today's quantum computers, whose limited number of qubits and noisy operations caused by hardware imperfections only permit a low number of data points or features (cf. \cite{preskill.2018a}).
The RQSVM model requires only $\mathcal{O}(\log_2(d))$ qubits for input data with $d$ features, making it particularly suitable for our application at hand.

To be precise, we consider \emph{milling}, a manufacturing process used in mechanical engineering, where a rotary cutter removes material from a workpiece to machine a desired product.
The resulting high-precision parts are essential for a wide range of industries such as aerospace~\cite{WS16}, automotive~\cite{DGK16} and medicine~\cite{FRD22}.
During milling, a number of factors, such as the spindle speed or wear state of the cutting tool, can affect process stability and cause location errors on the workpiece surface.
Predicting the stability of such processes accurately is critical for meeting the high industrial quality standards.

There is substantial work investigating stability in machining processes from a theoretical point of view~\cite{ASB+20} taking various physical properties of the spindle, cutting tool, machine tool and machined material into account.
This article takes a simplifying approach, assuming that we can model stability limits by means of a parametrized feature map and a least-squares regression model, sacrificing some accuracy for practicability.
Thereby, we are able to approach this problem with a quantum ML algorithm, demonstrating the feasibility of near-term quantum computing for real-world regression problems.
As the development of quantum computers progresses, this method may eventually exploit even richer feature maps~\cite{Schuld2019} and quantum speedup~\cite{nielsen.chuang.2010a}, leading to potential advantages over classical ML approaches.

Our contributions can be summarized as follows:
\begin{itemize}
	\item We propose a novel quantum ML regression model based on Support Vector Regression
	\item We formulate the problem of stability prediction in machining setups as a quantum ML problem
	\item We show that our method displays good performance in predicting the stability limit on real-world machining data sets
	\item We investigate the variation between milling cutters of the same make by predicting the stability limits of one tool from all others
\end{itemize}

\section{Background}
\label{sec:background}

\subsection{Stability of Machining Processes}
\label{sec:machining}

Milling processes are an essential part of manufacturing technology and are crucial for the precision machining of a wide range of materials, including metals, polymers and composites.
These processes use a superposition of translational and rotational movements of the milling cutters to remove material from a workpiece, enabling the production of complex shapes and high-precision components~\cite{din8589-3} using advanced techniques such as high-speed milling~\cite{tlusty93,SH95,ST97,KPK+20} and 5-axis milling~\cite{LBE11,HGB13}.
Such components are used in industries such as aerospace~\cite{WS16}, automotive~\cite{DGK16} and medical device manufacturing~\cite{FRD22}, where strict tolerances and complex shapes are often required.

Ensuring stable milling is an important aspect in process design, as unstable conditions can lead to \emph{chatter}, a self-excited vibration, which can significantly reduce the quality of the surface of the machined workpiece, lead to excessive tool wear, and potentially damage the machine tool~\cite{ASB+20}.
Chatter occurs mainly due to the dynamic interaction between the cutting tool and the workpiece, and can be influenced by various factors such as the spindle speed, depth of cut, and tool geometry.
Milling stability prediction commonly involves time-domain simulations~\cite{WSB+18,DPG+19}, frequency-domain analyses~\cite{MA04,FA22}, receptance coupling~\cite{SD00,SD05} or a combination of these approaches~\cite{ASM+08}.

However, various complex cause-effect relationships, such as concept drift~\cite{GZB+14} caused by tool wear, can lead to significant deviations between predicted and actual process stabilities.
This may lead to poor generalization of prediction strategies tailored to, e.g., a single machining center or tool.
In this case, data-driven approaches can be used to achieve comparably high prediction accuracy while generalizing across different scenarios of interest, such as the influence of multiple machine tools or tool wear on milling dynamics~\cite{WFS24}.

\subsection{Support Vector Machines}
\label{sec:svm}

The support vector machine (SVM) is a classification model that, in its original form, separates points belonging to one of two classes, $+1$ and $-1$ (see, e.g., \cite{hastie.etal.2009a}).
To this end, the SVM tries to find a hyperplane such that points of one class lie on one side, and points of the other class on the other side.
To achieve best possible generalization, the hyperplane is chosen such that it is as far away from the nearest points as possible, maximizing the \emph{margin} between points and decision boundary.
As perfect separability is unlikely, \emph{slack variables} $\xi_i$ allow for transgression of the separation property.
The SVM enjoys various appealing theoretical properties (cf., Vapnik et al.~\cite{Vapnik1998}), e.g., an upper bound for the generalization error of the SVM can be defined in terms of the width of its margin.
A classical SVM classifier solves the following optimization problem: \begin{equation}
	\begin{split}
		\min_{\w,b,\vec\xi}~ &\frac 1 2\w^\T\w+C\sum_{i=1}^\ell\xi_i \\
		\text{s.t.}~ &y_i(\w^\T\phi(\vec x)+b)\geq 1-\xi_i,\\
		&\xi_i\geq 0, ~\forall i\in\lbrace 1,\dots,\ell\rbrace.
	\end{split}
\end{equation}
Here, $y_i\in\lbrace +1,-1\rbrace$ is a binary class label, and $C>0$ is a hyperparameter that controls the impact of misclassified points, whose distance $\xi_i$ to the decision boundary gets penalized.
In addition, $\phi(\cdot)$ is a feature map that projects the original data into some higher-dimensional space.
Given the optimal $\w$ and $b$, the decision function of the trained Support Vector Classifier (SVC) model is \begin{equation}\label{eq:svc}
	f_{\text{SVC}}(\vec x;\w,b)=\sign(\w^\T\phi(\vec x)+b)\;,
\end{equation}
which is a binary indicator of which side of the decision boundary the new point $\vec x$ lies.

\subsection{Quantum Computing}
\label{sec:qc}

\emph{Quantum Computing} (QC) is a computing paradigm originating in the latter half of the 20\textsuperscript{th} century, which has gained widespread renewed attention due to the continuous improvement of physical quantum computers.
It holds the potential to solve certain computationally hard problems faster than any \emph{classical} (i.e., non-quantum) computer~\cite{shor1999polynomial,grover1996fast}, and, more recently, gave rise to \emph{quantum machine learning}, which aims to apply quantum computing techniques to machine learning tasks.

At the core of QC lies the idea to replace classical bits with \emph{quantum bits} (or \emph{qubits}, for short), and perform computations with them.
In contrast to bits, which can take one of two values, 0 or 1, qubits have two properties that go beyond.

\paragraph{Superposition}
Firstly, every qubit can be in a state that is neither 0 nor 1, but a mixture of both, called \emph{superposition}.
When \emph{measuring} a qubit $\ket{\psi}$ in superposition, it takes one of its two \emph{basis states} $\ket{0}$ or $\ket{1}$, with a certain probability determined by its state.
The state of the single qubit $\ket{\psi}$ can be described by a 2-dimensional complex-valued vector $\ket{\psi}=[\alpha_0, \alpha_1]^\T$ called \emph{amplitude vector}.
The notation $\ket{\psi}$ (say ``ket psi'') simply denotes a vector, with $\bra{\psi}$ (say ``bra phi'') its conjugate transpose $\bra{\psi}=(\ket{\psi})^\dagger=[\alpha_0^*,\alpha_1^*]$.
Each entry (or \emph{amplitude}) corresponds to a possible basis state, i.e., 0 and 1.
The absolute square of the amplitude yields the probability to be in the respective state~\cite{nielsen.chuang.2010a}.
For this reason, amplitude vectors always obey $\abs{\alpha_0}^2+\abs{\alpha_1}^2=1$, or simply $\braket{\psi}=1$ using the inner product.

The basis states corresponding to the classical bits 0 and 1 are $\ket{0}=[1,0]^\T$ and $\ket{0}=[0,1]^\T$.
However, a qubit with state $\ket{\psi}=\frac{1}{\sqrt{2}}[1,-i]^\T$ has an equal probability to be measured in either basis state, as \begin{equation*}\abs{\frac{1}{\sqrt{2}}}^2=\abs{\frac{-i}{\sqrt{2}}}^2=\frac{1}{2}\;.\end{equation*}
Note that, even though the amplitude for state 1 is negative and even complex, the measurement probability is the same, which shows that quantum states hold additional information which we cannot observe directly through measurement.

\paragraph{Entanglement}
A second special property of qubits is \emph{entanglement}:
If $n$ qubits $\ket{\psi^1},\dots,\ket{\psi^n}$ are in a system, their state is described by a combined amplitude vector $\ket{\Phi}=\ket{\psi^1\psi^2\dots\psi^n}=\ket{\psi^1}\otimes\ket{\psi^2}\otimes\dots\otimes\ket{\psi^n}$, where $\otimes$ is the \emph{Kronecker product}~\cite{nielsen.chuang.2010a}.
This vector has size $2^n$; again, each entry represents a basis state of the joint system, which is simply a binary string of length $n$, and also $\braket{\Phi}=1$.
However, through certain manipulations, we can construct quantum states that cannot be expressed as the Kronecker product of single qubits, such as the 2-qubit state $\ket{\Psi}=[1,0,0,1]^\T$ called \emph{Bell state}, which has an equal probability to be measured as \texttt{00} or \texttt{11}.
In this case, we say that the qubits are \emph{entangled}, which implies that their individual measurement probabilities are not independent.
Representing the measurement probabilities classically requires a table with $2^n$ entries, which is the reason why quantum computing is infeasible to simulate beyond a limited number of qubits.

\paragraph{Quantum Circuits}
A popular approach to quantum computing is through \emph{quantum circuits}, which are a graphical representation of sequential manipulations of a joint quantum state, reminiscent of logic circuits~\cite[Sec. 1.3.4]{nielsen.chuang.2010a}.

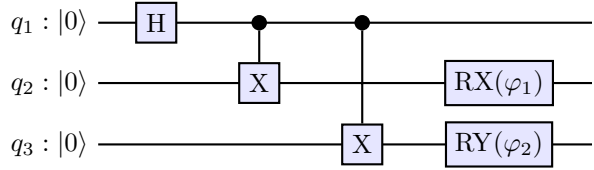
\begin{figure}
	\centering
	\begin{tikzpicture}[thick]
		\tikzstyle{gate} = [draw,fill=blue!10!white,minimum size=1.5em]
		\tikzstyle{ctrl} = [draw,fill,circle,minimum size=5pt,inner sep=0pt]
		\matrix[row sep=2mm, column sep=8mm] (circuit) {
			% qubit 1
			\node (q1) {$q_1:\ket{0}$}; &[-3mm]
			\node[gate] (H1) {H}; &
			\node[ctrl] (c1) {}; &
			\node[ctrl] (c2) {}; &&[-3mm]
			\coordinate (end1); \\
			% qubit 2
			\node (q2) {$q_2:\ket{0}$}; &&
			\node[gate] (x1) {X}; &&
			\node[gate] {$\mathrm{RX}(\varphi_1)$}; &
			\coordinate (end2); \\
			% qubit 3
			\node (q3) {$q_3:\ket{0}$}; &&&
			\node[gate] (x2) {X}; &
			\node[gate] {$\mathrm{RY}(\varphi_2)$}; &
			\coordinate (end3); \\
		};
		\begin{pgfonlayer}{background}
			\draw[thick]
				(q1) -- (end1) (q2) -- (end2) (q3) -- (end3) % qubits
				(c1) -- (x1) (c2) -- (x2); % controls
		\end{pgfonlayer}
	\end{tikzpicture}
	\caption{Exemplary quantum circuit with 3 qubits: All qubits are initialized in state $\ket{0}$. An H-gate is applied to $q_1$, putting it in superposition. Then, a controlled NOT-gate is applied to qubits $q_2$ and $q_3$, with $q_1$ serving as control. Finally, an RX gate is applied to $q_2$, and an RY gate to $q_3$, with parameters $\varphi_1$ and $\varphi_2$ respectively.}
	\label{fig:circuit}
\end{figure}

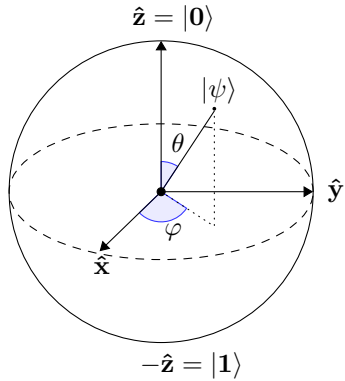
\begin{figure}
	\centering
	\begin{tikzpicture}[line cap=round, line join=round, >=Triangle]
		\clip(-2.19,-2.49) rectangle (2.66,2.58);
		\draw [shift={(0,0)}, blue!80!white, fill, fill opacity=0.1] (0,0) -- (56.7:0.4) arc (56.7:90.:0.4) -- cycle;
		\draw [shift={(0,0)}, blue!80!white, fill, fill opacity=0.1] (0,0) -- (-135.7:0.4) arc (-135.7:-33.2:0.4) -- cycle;
		\draw(0,0) circle (2cm);
		\draw [rotate around={0.:(0.,0.)},dash pattern=on 3pt off 3pt] (0,0) ellipse (2cm and 0.9cm);
		\draw (0,0)-- (0.70,1.07);
		\draw [->] (0,0) -- (0,2);
		\draw [->] (0,0) -- (-0.81,-0.79);
		\draw [->] (0,0) -- (2,0);
		\draw [dotted] (0.7,1)-- (0.7,-0.46);
		\draw [dotted] (0,0)-- (0.7,-0.46);
		\draw (-0.08,-0.3) node[anchor=north west] {$\varphi$};
		\draw (0.01,0.9) node[anchor=north west] {$\theta$};
		\draw (-1.01,-0.72) node[anchor=north west] {$\mathbf {\hat{x}}$};
		\draw (2.07,0.3) node[anchor=north west] {$\mathbf {\hat{y}}$};
		\draw (-0.5,2.6) node[anchor=north west] {$\mathbf {\hat{z}=|0\rangle}$};
		\draw (-0.4,-2) node[anchor=north west] {$-\mathbf {\hat{z}=|1\rangle}$};
		\draw (0.4,1.65) node[anchor=north west] {$\ket{\psi}$};
		\scriptsize
		\draw [fill] (0,0) circle (1.5pt);
		\draw [fill] (0.7,1.1) circle (0.5pt);
	\end{tikzpicture}
	\caption{The Bloch sphere, a 3d representation of a single-qubit quantum state; the basis states are at the north and south pole (image adapted from Nielsen and Chuang~\cite{nielsen.chuang.2010a}).}
	\label{fig:blochsphere}
\end{figure}

\Cref{fig:circuit} shows an example of a 3-qubit quantum circuit.
Operations are read from left to right:
Initially, all qubits are in state $\ket{0}$, yielding a joint state $\ket{000}=[1,0,0,\dots,0]^\T$.
An H-gate (\emph{Hadamard} gate) is applied to the first qubit, $q_1$.
Quantum gates represent linear transformations of the amplitude vector, i.e., complex-valued matrices of size $2^n\times 2^n$ that preserve the normalization property $\braket{\psi}=1$.
Such matrices $\mat U$ are called \emph{unitary}, with their defining property $\mat U^\dagger=\mat U^{-1}$.
As the entire joint state is always manipulated in its entirety at once, $q_2$ and $q_3$ remain unchanged, the operation carried out by the first gate is thus $\mat H\otimes\mat I\otimes\mat I$, where $\mat I$ is the 2d identity matrix.
The matrix $\mat H$ representing the Hadamard gate is defined as \begin{equation}
	\mat H = \frac{1}{\sqrt 2}\begin{bmatrix}1&1\\1&-1\end{bmatrix}\;,
\end{equation}%
and has the effect of taking $\ket{0}$ to an equal superposition $\mat H\ket{0}=\frac{1}{\sqrt 2}[1, 1]^\T$.
The following operation in \cref{fig:circuit} is a \emph{controlled NOT} (or \emph{CNOT}) gate, which inverts the state of the target qubit if the control qubit is 1, and leaves the target unchanged otherwise: \begin{equation}
	\CNOT=\begin{bmatrix}1&0&0&0\\0&1&0&0\\0&0&0&1\\0&0&1&0\end{bmatrix}.
\end{equation}%
The same is repeated with with qubit 3 as a target.
Finally, two \emph{rotation} gates RX and RY are applied to qubits 2 and 3 with parameters $\varphi_1$ and $\varphi_2$, respectively.
Intuitively, as amplitude vectors are normalized, we can view the state of a qubit as the surface of a unit sphere, the \emph{Bloch sphere} (see \cref{fig:blochsphere});
applying RX and RY has the effect of rotating this sphere around the X and Y axis by a certain angle, hence $\varphi_1,\varphi_2\in[0,2\pi]$~\cite[Sec. 4.2]{nielsen.chuang.2010a}.

Rotation gates like RX, RY and RZ are ways to input classical data into a quantum circuit, which is the foundation of variational quantum circuits (VQC).
Taking repeated measurements of the final state of the circuit (i.e., sampling from the resulting probability distribution) and counting the observed basis states serve as data output, e.g., by computing the empirical Bernoulli probability of individual qubits or approximating the overall joint distribution.
How this information is interpreted depends on the respective application.
Naturally, the more measurements are taken, the less sampling noise is present in the output.
However, the entire circuit has to be executed again for each measurement, because a measurement destroys the quantum state.

How to find a suitable and efficient circuit for a given application (e.g., a circuit that performs a specific operation on the entire quantum state or that produces a state that optimizes some criterion) is subject to ongoing research.
Some heuristic approaches vary the parameters of pre-defined multi-purpose circuits (``ans\"atze'')~\cite{peruzzo.etal.2014a}, while others construct entire circuits through iterative optimization~\cite{franken.etal.2022a}.
Other circuits, like the RQSVM described in \cref{sec:qsvm}, can be derived mathematically from the problem statement, which yields more theoretically sound quantum algorithms.

To summarize this section: \begin{itemize}
	\item Qubits have internal states that determine their probabilities of being measured as \texttt{0} or \texttt{1};
	\item Multiple qubits form systems that can be entangled, such that the measurement probabilities are not statistically independent;
	\item Quantum circuits perform linear operations on quantum states, altering the resulting probability distribution;
	\item Data can be injected into a circuit through angle parameters in rotation gates;
	\item Output data is computed from repeated measurements of the final quantum state after applying the circuit.
\end{itemize}

\section{Quantum Support Vector Regression}
\label{sec:qsvm}

Using the building blocks presented in the previous section, we constructed a quantum regression model based on an SVM~\cite{Cortes1995}.
To this end, we used the recently proposed Real-Part Quantum SVM (RQSVM)~\cite{piatkowski.muecke.2024a} as a basis and modified it to perform regression instead of classification.
We used $\epsilon$-Support Vector Regression~\cite{Chang2011,Vapnik1998} as the underlying classical model, whose fitted parameters $\w$ we embedded into a quantum circuit and performed regression by repeated measurement, allowing us to compute the inner product between $\w$ and our data features $\vec\phi(\vec x)$ and add a bias $b$, which yielded our prediction $\hat y$, \begin{equation}\label{eq:svr}
	f(\vec x;\w,b)=\w^\T\phi(\vec x)+b=\hat y\;.
\end{equation}

The RQSVM encodes the parameter vector $\w$ of an SVM model and an input data point $\vec x$ as a quantum circuit $C_{\w}(\vec x)$, whose resulting quantum state $\ket{\psi_{\w}(\vec x)}$ can be measured to approximate the inner product $\w^{\T}\phi(\vec x)$.
To construct this circuit, several components were required, which are described in the following.

\paragraph{Unitary Vector Embedding}
Given a vector $\vec v\in[-1,1]^d$ with $d>1$, let $n=\lceil \log_2(d) \rceil$, where $\lceil\cdot\rceil$ denotes rounding up to the nearest integer.
The $2^{n}\times 2^{n}$-matrix $\Delta(\vec v)$ whose elements are given by \begin{equation}
	\Delta(v)_{j,k} = \begin{cases}
		\exp(-i\arccos(v_j))&\text{if }j=k \text{ and } j\leq d,\\
		-i&\text{if }j=k \text{ and } j > d,\\
		0&\text{otherwise}
	\end{cases}
\end{equation}
is diagonal and unitary.
This technique allowed us to embed arbitrary bounded vectors into a unitary matrix.

\paragraph{Real-Part Extractor}
Given a diagonal unitary matrix $S$, the unitary matrix \begin{equation}
	R(\mat S)= (\mat H\otimes \mat I^{\otimes n}) (\ketbra{0}\otimes\mat S+\ketbra{1}\otimes \mat S^\dagger) (\mat H\otimes \mat I^{\otimes n})
\end{equation}%
allowed us to apply $\Re(\mat S)$, i.e., the real part of matrix $\mat S$, on some arbitrary quantum state $\ket{\psi}$.
To do this, we executed $R(\mat S)(\ket{0}\otimes\ket{\psi})$, and if we measured the first qubit as $\ket 0$, the circuit successfully executed the non-unitary operation $\frac 1 2(\mat S+\mat S^{\dagger})\ket\psi=(\Re(\mat S))\ket\psi$.

\paragraph{Sign Expansion}
Given a vector $\vec v\in[-1,1]^d$ for some $d>1$, the vector \begin{equation}
\vec v_{\pm}= (\ket{0}\otimes\vec v_-)+(\ket{1}\otimes\vec v_+),
\end{equation}
where $\vec v_+$ ($\vec v_-$) replaces all negative (positive) entries of $\vec v$ by $0$, is in $[0,1]^{2d}$, effectively creating a ``sign qubit'': \begin{align*}
	\vec v_+ &= (\max\lbrace 0,v_1\rbrace,\dots,\max\lbrace 0,v_d\rbrace)^\T, \\
	\vec v_- &= (-\min\lbrace 0,v_1\rbrace,\dots,-\min\lbrace 0,v_d\rbrace)^\T.
\end{align*}

\paragraph{The Real-Part Quantum SVM}
Finally, we put these components together to obtain \begin{align}
	C_{\w}(\vec x) &=R(W(\w/\norm{\w}_{\infty})U(\vec x)),\\
	\text{where } W(\w)&=\ketbra{0}\otimes\Delta(\sqrt{\w_{\pm}})+\ketbra{1}\otimes\Delta(\sqrt{\w_{\pm}})^{\dagger},\nonumber\\
	U(\vec x)&=R(\Delta(\sqrt{\phi(\vec x)_{\pm}/\norm{\phi(\vec x)}_{\infty}})).\nonumber
\end{align}
If we prepare an initial quantum state $\ket{\psi_0}=\ket{0}^{\otimes 3}\otimes(\mat H\otimes\ket 0)^{\otimes m}$ with $m=\lceil\log_2(d)\rceil$, the resulting state $\ket{\psi_{\w,\vec x}}:=C_{\w}(\vec x)\ket{\psi_0}$ has measurement probabilities of the form $c\cdot \abs{w_j}\phi(\vec x)_j$, where $c$ is a constant.
As we applied sign expansion to $\w$, we could reconstruct the original signed values despite the fact that probabilities are always non-negative.
This allowed us to compute the inner product $\w^{\T}\phi(\vec x)$ by repeatedly measuring $\ket{\psi_{\w,\vec x}}$.
Finally, the model output was the sign of the inner product plus the bias $b$, as in \cref{eq:svc}.

%\medskip
Extending the RQSVM to a regression model involved swapping out the training procedure from an SVM classifier to an $\epsilon$-SVR, using the resulting weight vector $\w$ for $C_{\w}$, and omitting the sign function for the output.

\section{Application}

In a series of experiments, we applied our RQSVR model for predicting the stability limits in our data sets $\mathcal D_A$ and $\mathcal D_B$.
To this end, we followed a two-step approach: \begin{enumerate*}[label=(\roman*)]
	\item We defined a suitable feature map $\vec\phi$ that was able to capture the data behavior, and then
	\item trained RQSVR models on the resulting features in order to predict the stability limit.
\end{enumerate*} First, we give an overview of the machining data used throughout this article.

\subsection{Data}
\label{sec:data}

\begin{table}
	\centering
	\caption{Overview of our data collection setup.}
	\label{tab:datacollection}
	\begin{tabular}{ll}
		\toprule
		Tool diameter &$d=\SI{12}{mm}$ \\
		Number of flutes &$4$\\
		Projection length &$\SI{48}{mm}$\\
		Workpiece &Steel AISI 4140, soft annealed\\
		Spindle speed &$n\in[4000, 8000]\,\text{RPM}$\\
		&$\Delta n=\SI{50}{RPM}$\\
		Axial depth of cut &$a_p=\SI{4.6}{mm}$\\
		Tooth feed &$f_z=\SI{0.08}{mm}$\\
		Milling strategy &Side milling with linearly increasing $a_e$\\
		Machining center &$\text{DMU}_A$: DMU 50\\
		&$\text{DMU}_B$: DMU 50 eVolution\\
		\bottomrule
	\end{tabular}
\end{table}

\begin{table}
	\centering
	\caption{Description of the columns of our machining data set.}
	\label{tab:dataset}
	\begin{tabular}{rll}
		\toprule
		Name &Domain &Description \\ \midrule
		$\xa$ &$\lbrace 4000,\dots,8000\rbrace$ &rotation speed in RPM\\
        $\xb$ &$[0, 263.725]$ &tool wear condition in \si{\centi\meter\squared}\\
        $\y$ &$[1.08, 5.44]$ &stability limit in \si{\milli\meter}\\
		\bottomrule
	\end{tabular}
\end{table}

The data set described in the following was also used in a study considering the transfer of tool wear-dependent stability predictions to multiple machine tools based on classical ML methods in order to reduce the required experimental efforts for data acquisition~\cite{WFS24}.
A detailed description of the data collection setup can be found in the \hyperref[sec:datacollection]{appendix}.
An overview is shown in \cref{tab:datacollection}.

The aim of the experiments was to evaluate the dynamic behavior of the milling process under varying spindle speed $\xa$ and tool wear $\xb$.
We measured acoustic emission signals acquired during milling while increasing the radial cutting depth $a_e$ at each spindle speed $\xa$ and analyzed them to quantify the chatter intensity.
The tests were carried out with two machining centers, DMU 50 ($\text{DMU}_A$) and DMU 50 eVolution ($\text{DMU}_B$), resulting in two data sets, $\mathcal D_A$ and $\mathcal D_B$.
By using two different machining centers we increase the generality of our data, allowing for comparisons across different setups.
In total, 1037 and 1065 milling tests were carried out on $\text{DMU}_A$ and $\text{DMU}_B$, respectively.
For this purpose, six milling tools were used, labeled $T_A^{(1)}$, $T_A^{(2)}$ and $T_A^{(3)}$ for $\mathcal{D}_A$, and $T_B^{(1)}$, $T_B^{(2)}$ and $T_B^{(3)}$ for $\mathcal{D}_B$.
Unless specified otherwise, we simply combine all measurements of the individual tools within each data set, as they should (theoretically) display the same behavior.
In fact, we test this hypothesis in \cref{sec:tools}.

\begin{figure}
	\centering
	\includegraphics{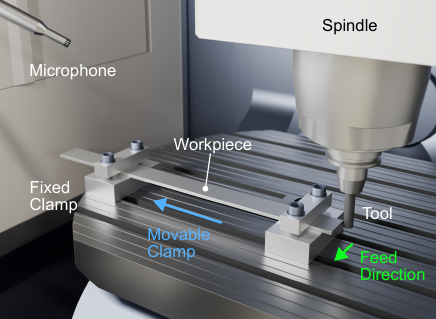}
    \caption{Schematic visualization of the experimental setup for the milling investigations~\cite{WFS24}.
    }
	\label{fig:exp}
\end{figure}

\cref{fig:exp} shows a schematic visualization of the experimental setup used for the milling processes.
Stability limits, which serve as targets $\y$, were determined by applying thresholds to the accumulated chatter intensities in the frequency range of \SIrange{500}{8000}{\hertz}, using a spectrogram window size of \SI{0.01}{\second}.
Some processes were entirely stable, i.e., $\y>a_{e,\text{max}}$, and we dropped them from our data sets, resulting in $895$ usable data points for $\text{DMU}_A$ and $829$ data points for $\text{DMU}_B$.
The two primary features $\xa$ and $\xb$ are collected in a vector $\vec x=(\xa, \xb)^\T$.
The entire data sets collected from $\text{DMU}_A$ and $\text{DMU}_B$ are denoted as $\mathcal{D}_A$ and $\mathcal{D}_B$, each containing pairs $(\vec x,\y)$ of feature vector $\vec x$ containing spindle speed and tool wear and the resulting stability limit $\y$, with $\abs{\mathcal{D}_A}=895$ and $\abs{\mathcal{D}_B}=829$.
For an overview of the data, see \cref{tab:dataset}.

\subsection{Feature Extraction}
\label{sec:feat}

\begin{figure*}
	\centering
	\includegraphics[width=\textwidth]{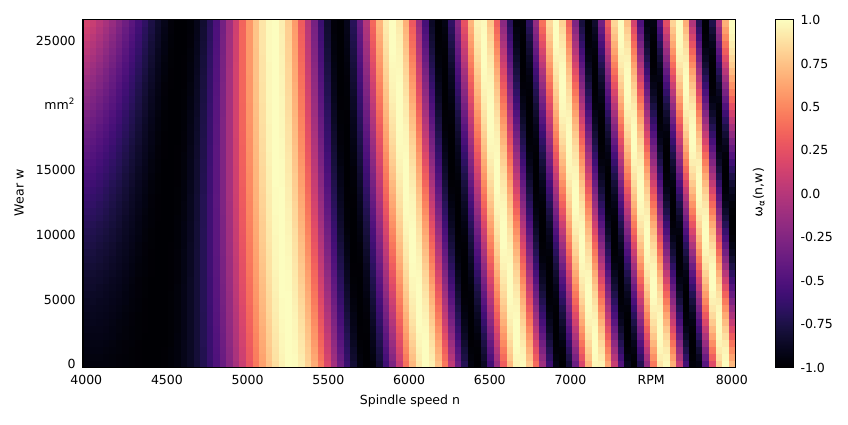}
	\caption{Example of the function $\omega_{\vec\alpha}$ used in the feature map $\vec\phi_{\cos}$, containing a $\cos$ term to capture the wave-like pattern exhibited by the stability limits.}
	\label{fig:wave}
\end{figure*}

Since milling dynamics were considered, the data contained non-linearities.
For this reason, we constructed a feature map $\vec\phi_{\cos}$ that contained a cosine term in order to capture the wave-like behavior resulting from these non-linearities.
However, we used a degree-2 polynomial over $\xa$ and $\xb$ as the argument of $\cos$, allowing for waves with varying instead of constant frequency.
We defined the feature map as \begin{multline}
	\vec\phi_{\cos}(\vec x;\vec\alpha)=\bigl[\xa,\xb,\xa^2,\xa\xb,\xb^2,\omega_{\vec\alpha}(\vec x),\xa\omega_{\vec\alpha}(\vec x),\xb\omega_{\vec\alpha}(\vec x),\xa^2\omega_{\vec\alpha}(\vec x),\xa\xb\omega_{\vec\alpha}(\vec x),\xb^2\omega_{\vec\alpha}(\vec x)\bigr]^\T,
\end{multline}
with \begin{equation*}
	\omega_{\vec\alpha}(\vec x)=\cos(\alpha_1+\alpha_2\xa+\alpha_3\xb+\alpha_4\xa^2+\alpha_5\xa\xb+\alpha_6\xb^2).
\end{equation*}
The function $\omega_{\vec\alpha}$ is shown exemplarily in \cref{fig:wave}.

In the following, we write $\mathcal D$, however, we performed each step separately on $\mathcal D_A$ and $\mathcal D_B$.
We found the values $\vec\alpha=(\alpha_1,\dots,\alpha_6)$ by defining \begin{equation}f(\vec x;\vec\alpha,\vec\beta,c)=\vec\beta^{\T}\vec\phi_{\cos}(\vec x;\vec\alpha)+c\end{equation} and performing a least-squares fit on the normalized data set.
To this end, we transformed $x_i\mapsto (x_i-m_i)/s_i$, where $m_i=\min_{x\in\mathcal D}x_i$ and $s_i=(\max_{x\in\mathcal D}x_i)-m_i$, for $i\in\lbrace 1,2\rbrace$.
Then we used the L-BFGS-B algorithm~\cite{liu1989limited} to minimize the mean squared error (MSE) of $f$ w.r.t. $\mathcal{D}$, using the implementation provided by the \texttt{scipy} Python package\footnote{\url{https://scipy.org}}.
To search the parameter space more thoroughly and obtain better results, we sampled the starting point $\bm\alpha^{(0)}$ for the optimization algorithm from $\mathcal{N}(\bm\mu=(0,\dots,0)^\T,\bm\Sigma=10\bm I)$ (each component i.i.d. normally distributed around $0$ with standard deviation $10$) and performed a simple local search by sampling $\bm\epsilon^{(t)}\sim\mathcal N(\bm\mu=(0,\dots,0)^\T,\bm\Sigma_t=10\bm I)$ and setting $\bm\alpha^{(t+1)}=\bm\alpha^{(t)}_{\text{min}}+\bm\epsilon^{(t)}$, where $\bm\alpha^{(t)}_{\text{min}}$ is the starting point that yielded the lowest MSE after running the L-BFGS-B algorithm from this starting point, up to and including step $t$, for $t>0$.
We followed this procedure until we observed no improvement after 1000 steps.

The final two models achieved an MSE of $0.12337479$ on $\mathcal D_A$ and $0.21384462$ on $\mathcal D_B$.
The $\alpha$ values we found and consequently used for $\omega_{\vec\alpha}(\vec\alpha)$ are listed in \cref{tab:parameters}.

\begin{table}
	\centering
	\caption{Parameters $\bm\alpha$ found for $\mathcal D_A$ and $\mathcal D_B$.}
	\label{tab:parameters}
	\begin{tabular}{lrr}
		%\hline
		Parameter &$\mathcal D_A$ &$\mathcal D_B$\\ \hline
		$\alpha_1$ &$15.52749483$  &$-2.87183284$ \\
		$\alpha_2$ &$-4.46971848$  &$11.95336368$ \\
		$\alpha_3$ &$-9.55499409$  &$0.99781459$  \\
		$\alpha_4$ &$-14.22448621$ &$28.58595761$ \\
		$\alpha_5$ &$-2.38118747$  &$2.27955561$ \\
		$\alpha_6$ &$14.75707122$  &$-4.34204599$
		%\hline
	\end{tabular}
\end{table}

Finally, we transformed all data points in $\mathcal{D}$ and defined the new data set \begin{equation}
	\mathcal{F}=\left\{(\vec\phi_{\cos}\left((\vec x^{\ell}-\vec m)/\vec s;\vec\alpha\right),\y^{\ell}): (\vec x^{\ell},\y^{\ell})\in\mathcal{D}\right\},
\end{equation}
with $\vec m=(m_1,m_2)^\T$ and $\vec s=(s_1,s_2)^\T$.
This yielded $\mathcal F_A$ and $\mathcal F_B$ consisting of 11 features each.

\subsection{RQSVR Training}

We used $\mathcal F$ to train a RQSVR model as described in \cref{sec:qsvm}.
For training the underlying $\epsilon$-SVR models we used the implementation from scikit-learn~\cite{scikit-learn}.
The final RQSVR model was simulated using a custom quantum simulator implemented in Python using NumPy~\cite{harris2020array}.

As a first step, we determined the best value for the hyperparameter $C$, for which we performed a grid search with values $C=10^k$ for $k=-3,-2,\dots,2$.
For each choice of $C$, we performed a 10-fold cross validation by shuffling the data, splitting it into 10 subsets, and predicting each subset after training on the remaining 9 for each subset in turn.
This yielded ten models, for each of which we recorded the MSE.
To compare the hyperparameter settings, we used the mean over all ten MSE values.
For better comparability, we used the same data set splits across all hyperparameter values.

\begin{figure*}
	\centering
	\includegraphics[width=\textwidth]{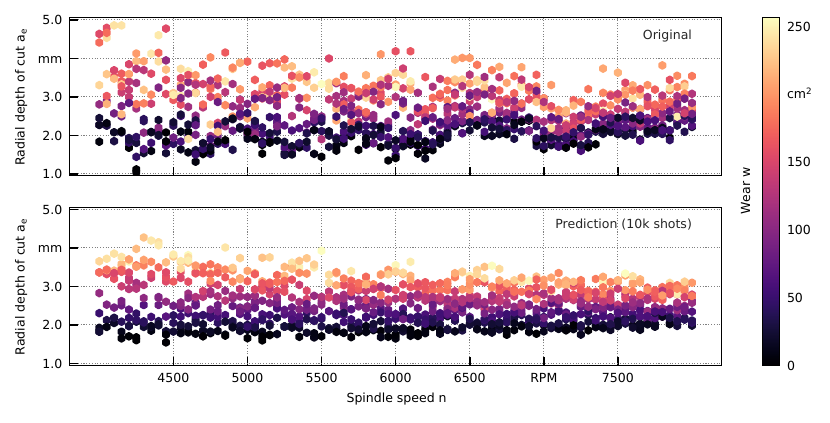}
	\caption{\textbf{Top:} Original data set $\mathcal{D}_A$ obtained using machining center $\text{DMU}_A$.
		\textbf{Bottom:} Data set where the stability limit (y-axis) was predicted using the RQSVR model trained on the eleven features found in \cref{sec:feat}.}
\label{fig:altedmu_prediction}
\end{figure*}

\begin{figure*}
	\centering
	\includegraphics[width=\textwidth]{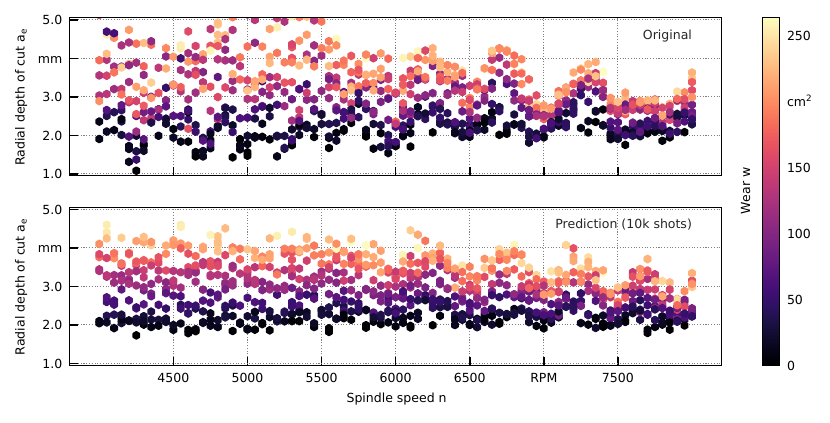}
	\caption{\textbf{Top:} Original data set $\mathcal{D}_B$ obtained from machining center $\text{DMU}_B$.
		\textbf{Bottom:} Data set where the stability limit (y-axis) was predicted using the RQSVR model trained on the eleven features found in \cref{sec:feat}.}
		\label{fig:prediction}
\end{figure*}

\begin{figure*}
	\centering
	\includegraphics[width=\textwidth]{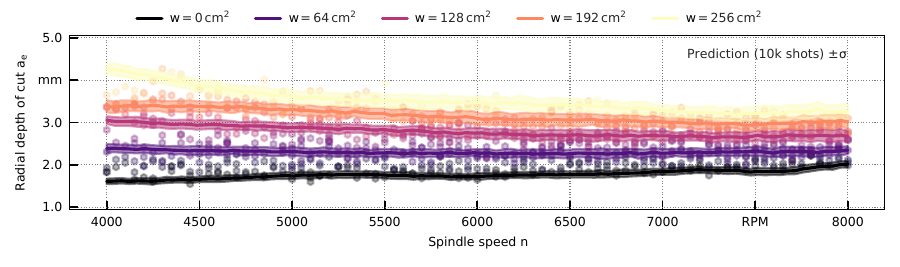}
	\caption{Model output for $\mathcal D_A$ for varying fixed wear values, with additional uncertainty regions showing one standard deviation over shot noise.}
	\label{fig:altedmu_modelfunction}
\end{figure*}

\begin{figure*}
	\centering
	\includegraphics[width=\textwidth]{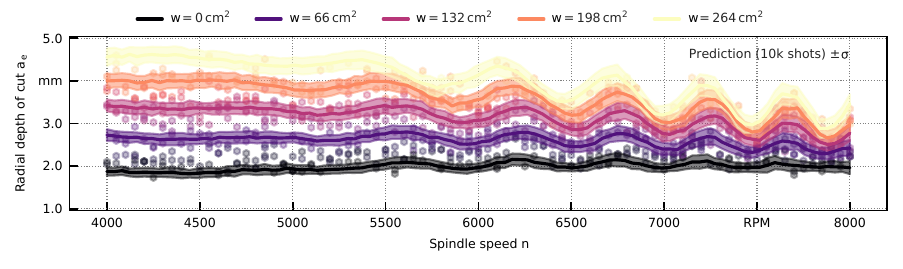}
	\caption{Model output for $\mathcal D_B$ for varying fixed wear values, with additional uncertainty regions showing one standard deviation over shot noise.}
	\label{fig:modelfunction}
\end{figure*}

In our experiment, we found the value $C=1$ to perform best, leading to a mean MSE of $0.15470591$ on $\mathcal D_A$ and $0.25276193$ on $\mathcal D_B$, using their respective feature maps.
\Cref{fig:altedmu_prediction,fig:prediction} shows the original data as well as the predictions of the training data each computed from 10k shots of our quantum simulator.
\Cref{fig:altedmu_modelfunction,fig:modelfunction} shows the model prediction at five different fixed levels of tool wear $\xb$ with areas of uncertainty, showing the empirical standard deviation caused by shot noise over 100 runs, again using 10k shots per prediction.

Using the feature map $\vec\phi_{\cos}$ described in \cref{sec:feat}, the model was able to capture the wave-like pattern present in the original data, including the smaller amplitude with lower rotation speed.
Generally, the model trained on $\mathcal{D}_B$ was more accurate, following the wave-like pattern more closely, while the model for $\mathcal{D}_A$ uses the cosine component only marginally and with much lower frequencies.
We suspect there might be another local optimum in the parameter space of $\bm\alpha$ which may be able to model the data behavior better.
For $\mathcal D_B$ and a rotation speed of about \SI{7700}{RPM}, the model predicted another peak, which was, however, not present in the input data -- here, the feature map seems to lack expressivity to capture this particular data behavior.
However, up to around \SI{7500}{RPM}, the visual performance of the model was high.

\subsection{Tool Prediction}
\label{sec:tools}

\begin{figure}[t]
	\centering
	\includegraphics[width=.6\columnwidth]{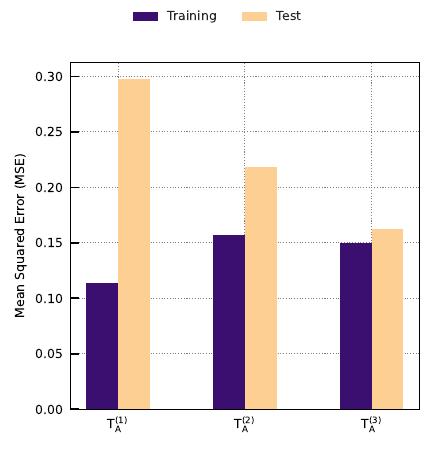}
	\caption{Mean squared error (MSE) of RQSVR models trained on tools in $\mathcal D_A$ (lower is better):
        For each tool ($T_A^{(1)}$, $T_A^{(2)}$ and $T_A^{(3)}$) one model was trained on the other two tools.
		To this end, the feature extraction described in \cref{sec:feat} was performed on only the other two tool, and then the RQSVR was trained using the resulting features.
		The quantum model used 10,000 shots for each prediction.}
	\label{fig:altedmu_toolmse}
\end{figure}

\begin{figure}[t]
	\centering
	\includegraphics[width=.6\columnwidth]{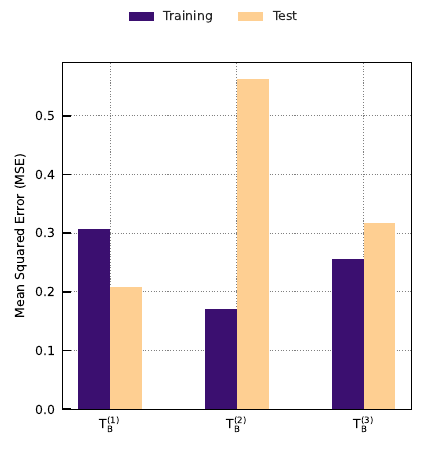}
	\caption{Mean squared error (MSE) of RQSVR models trained on $\mathcal D_B$ (lower is better):
		Same as in \cref{fig:altedmu_toolmse}, but for tools $T_B^{(1)}$, $T_B^{(2)}$ and $T_B^{(3)}$.}
	\label{fig:toolmse}
\end{figure}

In an additional experiment, we investigated how similar the tools contained in the data sets ($T_A^{(1)}$, $T_A^{(2)}$ and $T_A^{(3)}$ for $\mathcal D_A$, and $T_B^{(1)}$, $T_B^{(2)}$ and $T_B^{(3)}$ for $\mathcal D_B$) behaved. 
While in theory, the tools had the same size and should behave identically during the milling processes, due to imperfections they displayed slightly different stability limits.
This raised the question of how well we can predict the stability properties of one tool knowing the properties of all other tools of the tool type.
Ideally, we should have observed that each tool can be predicted equally well from the other tools, with slight variations introduced by the measuring process.

We performed our study separately on $\mathcal D_A$ and $\mathcal D_B$.
For each of the three tools, we split the data into a training and a test set, such that the training set contained two tools, and the test set the third.
Then we performed feature extraction as described in \cref{sec:feat} on only the training set, obtaining features that are optimized toward predicting the stability limit of two tools.
Using these features, we trained an RQSVR model, using hyperparameters $\epsilon=0.1$ and $C=1$.

The results can be seen in \cref{fig:altedmu_toolmse,fig:toolmse}.
Interestingly, both data sets contained one tool which seemed to deviate from the other two
In $\mathcal D_A$, tool $T_A^{(1)}$ yielded a comparatively high test MSE, indicating that its stability limits were harder to predict from the other two tools.
As for $\mathcal D_B$, $T_B^{(2)}$ seemed to behave different from the other tools.
While the training error, i.e., after training on both $T_B^{(1)}$ and $T_B^{(3)}$, was lowest, the test error was about twice as high as for the other combinations, implying that $T_B^{(2)}$ could not be predicted very precisely from the other tools.
In contrast, tools $T_B^{(1)}$ and $T_B^{(3)}$ yielded relatively low test MSE of around $0.2$ and $0.3$, respectively.
Similarly, tools $T_A^{(2)}$ and $T_A^{(3)}$ were in a range of test MSE values between $0.15$ and $0.22$.
%The ``outlier tool'' WZ9 in $\mathcal D_A$ is less pronounced than 
Finally, the test error of the $T_B^{(1)}$ model was lower than its training error, which is unusual.
We guess that the prediction of $T_B^{(2)}$ added so much error to the training error that it surpassed the test error.

\section{Conclusion}

In this work we have studied the use of quantum-based regression models for predicting stability limits of milling tools in machining processes.
To this end, we derived a feature map from the stability limits that we found experimentally on a series of identical milling tools on two different machining centers by varying the spindle speed and the wear condition of each tool and identifying the radial depth of cut where chatter occurred.
The resulting features could be used as input to a Real-part Quantum SVM, which predicted the stability limits.
We found that we could model this limit to satisfying accuracy using our proposed method.

Our contributions are twofold:
We extended the RQSVM model, which is originally a classifier, to perform regression.
This novel Real-part Quantum Support Vector Regressor (RQSVR) is another step in the direction of quantum-ready Machine Learning models; just like the RQSVM, the RQSVR preserves the theoretical properties of its classical counterpart, the SVR model, adding only sampling noise introduced by quantum measurement.
This preservation of theoretical guarantees sets this model apart from other heuristic methods commonly found throughout Quantum ML, which fall into completely different model classes.

Moreover, as an exemplary real-world application, we showed that the stability of machining processes could be modeled and predicted through ML models that can be deployed on quantum devices, ultimately extending the toolbox of available methods to improve milling quality.
As quantum computing hardware keeps improving, we expect our method to become a viable alternative to ML on classical hardware:
Quantum computers beyond the Noisy Intermediate-Scale Quantum era~\cite{preskill.2018a} will be capable of computing rich feature maps with high fidelity (see, e.g., \cite{schuld2021quantum}), allowing for faster and more accurate classification of stability limits based on our presented method.

In addition, we investigated the variation between milling cutters of the same type by trying to predict the stability limits of one tool from a model trained on all others.
This experiment demonstrated that milling tools, despite being of the same size and type, exhibit slightly varying properties that can be detected using our quantum model.

For future work, we will deploy our method on real quantum hardware as soon as it is sufficiently noise-free and allows for sufficiently deep circuits to be executed.
Continuing the path towards theoretically sound QML models, discovering and applying quantum counterparts of more ML models (e.g., Artificial Neural Networks or Random Forests) that preserve their theoretical properties seems highly promising.

With regard to quantum computing, discovering a quantum feature map that captures the lobes of stability which \cref{fig:altedmu_prediction,fig:prediction} exhibit could further increase the accuracy of our quantum model significantly.
As shown by Schuld et al.~\cite{schuld2021effect}, certain data embeddings create quantum feature spaces consisting of truncated Fourier series, which can encode periodic signals.
Tying such feature maps to theoretical work in this area is a promising open problem.

\clearpage
\appendix
\section{Data Collection Setup}
\label{sec:datacollection}

For the milling tests, four-flute end mills from Seco Tools with a diameter of \SI{12}{\milli\meter} were used.
AISI 4140 steel in the soft annealed condition was used as the workpiece material.
The aim of the tests was to evaluate the dynamic behavior of the milling process with regard to tool wear.
This was achieved by using several milling tools under different wear conditions.
As a result, both the spindle speed $\xa$ and the tool wear condition $\xb$, which was defined in proportion to the amount of material removed per unit length of the tool used, were varied in the tests.
For the spindle speed, a range of \SIrange{4000}{8000}{RPM} was used in order to comply for the specifications of the workpiece material and cutter diameter used.
A spindle speed increment of \SI{50}{RPM} resulted in 81 different spindle speed values.
The axial depth of cut was held constant at $a_p = \SI{4.6}{\milli\meter}$.
A tooth feed of $f_z = \SI{0.08}{\milli\meter}$ resulted in varying feed velocities $v_f$ depending on the selected spindle speed.
Different levels of tool wear were generated by using a single tool across multiple experiments, allowing the wear to progressively increase with each cutting test.
Since this approach would yield a data set where each $\xb$ value appeared only once, three new tools were employed until a wear state of approximately \SI{250}{\centi\meter\squared} was reached.
The radial cutting depth $a_e$ was varied by side milling with linearly increasing $a_e$ up to $a_{e,\text{max}}$, which was selected on the basis of the current wear condition and the expected stability behavior.

The tests were carried out with two machining centers, DMU 50 ($\text{DMU}_A$) and DMU 50 eVolution ($\text{DMU}_B$).
A microphone sensor (PCB-130F20) was used to determine the process stability.
The same tool types and tool holders were used on both machines, whereby a fixed projection length of \SI{48}{\milli\meter} was maintained.
A total number of 1037 and 1065 milling tests were carried out on $\text{DMU}_A$ and $\text{DMU}_B$, respectively.
\end{document}